\documentclass[preprintnumbers,nofootinbib,noshowpacs,eqsecnum,twocolumn,prd]{revtex4}

\usepackage{url}
\usepackage{graphicx}
\usepackage{array,color}

\begin{document}

\title{Illusions - a model of mind}

\author{Markos Maniatis}
\affiliation{UBB, Departamento de Ciencias Basicas, Chillan, Chile}
%

\begin{abstract}
Recognizing that all mental processes have to be unfree and passive, we
develop a model of
behavior and perceptions. We shall see how misleading our 
intuition is and
shall understand how consciousness arises. 
\end{abstract}


\maketitle


\section{Introduction}
\label{einleitung}


We are convinced to be like the captain of a ship - equipped 
with navigation systems like radar, depth sounder on the one hand
und rudder, radio equipment, switches and levers for valves and locks
on the other hand. 
Similar, we get on the one hand visual, auditive, haptic and other
sensory informations 
and on the other hand we control and steer our body, 
walk, grasp, gesticulate und communicate.
We are convinced to be {\em free} in the sense, 
to indeed have informations available about our surrounding and our body, but at least
expect to be able to control ourselves to a certain level arbitrary. Of course,
we know that certain processes are unconscious, like, for instance,
the control of our heart. However, we expect to be free at least
with respect to {\em conscious} behavior.

This picture of a ''captain'' on board of our body reveals at a closer look
as an illusion; on general grounds it appears to be meaningless to 
present the sensory signals to any kind of inner ''captain''; our
senses have accomplished this task already in transferring the
perceptions to our nervous system, that is, have made the information 
accessible for further processing.
Why should these informations be presented once again to an inner
''captain'' \cite{Rosenthal86}? In particular, the informations would have to be 
processed once again within the ''captain'' and we were no step further. 
In fact, the incoming signals, for instance of the visual system, are
processed already behind the retina and are directed to different regions
in the cortex. Accordingly, there is no localisation in our brain in which the
visual signals converge.
Obviously, we have to abandon the illusion of a ''captain''. But how
does this illusion arise?

A further illusion of the ''captain'' is his {\em freedom}: 
we think of a ''captain'', who has information
available for instance on monitors, about the current position of the ship
and its velocity, but we imagine that the ''captain'' is in principle {\em free}.
We expect the ''captain'' to balance different options and to have
a certain range of possibilities to decide and act. 
In this sense we speak of a kind of responsibility of the ''captain''. 
We will see that also this {\em freedom} of the ''captain'' is an illusion
when we consider the ''captain'' in our nervous system. 

The absence of our freedom is the
crucial point in order to understand our
behavior, thinking, our perceptions, and eventually
consciousness. The quest for free will is certainly 
very old \cite{Democritus, Aristotle}
- and it is still subject to discussions today.
In section \ref{willensunfreiheit} we shall discuss in detail this
question about our freedom.

 
Historically it seemed to be obvious that we are
not free, since two break thoughts have been achieved in science:
firstly, it became clear that physiological processes are 
in principle not different from other natural phenomena \cite{Woehler}. 
Secondly, the principle of cause and effect, the determinism,
was recognized as a fundamental and universal principle. These
two findings lead to the following question:
how could we be free, if our physiological processes have
to follow the principle of cause and effect, that is, 
are deterministic processes?
We want to discuss some aspects of this discussion in the section \ref{willensunfreiheit} and
we will argue that our freedom is an illusion, in a certain sense
independent of the question of determinism.

If we are not free, the question arises, how do we come
to our decisions and actions? 
Since the free ''captain'' in us turns out to be an illusion, the question is,
how do we steer our ship without any type of ''captain''?
Obviously it appears that we in general do not behave like a ghost ship -
we act in general purposively. As a consequence of the absence of freedom
it is required to replace the ''captain'' by a non-free, that is, passive ''mechanism'' 
in order to understand our decisions and actions consistently. 
This ''mechanism'' has to connect the incoming signals from our 
perceptions, which arrive at
our nervous system with the outgoing ones, which represent our decisions and actions. 
In section \ref{handeln} we shall present the ''mechanism'' which allows us
to replace the ''captain'' and shall give a consistent explanation of our 
behavior and thinking.


In section \ref{wahrnehmung} we will focus on our perceptions. 
If we, for instance, watch the sunset
our sensation seems not to be in accordance
with electrical action potential in the neurons of our nervous system. 
With other words, the question arises, how does the electric action-potential activity
of neurons correspond to the sensation of watching a sunset? What is the meaning
of pain when we cut accidentally our skin with a knife in contrast to
the electrical or biochemical neuronal activity?  
We recognize the problem already when we ask the simple question of the
sensation of a color, say red: the color red corresponds physically to a range
of wave length of an electromagnetic wave. This electromagnetic wave excites 
charges to oscillate in special cells in our retina which in turn induce 
electrical actions potentials in the neurons. The signals are transferred to the cortex
and the whole processing in our nervous system is performed 
in form of action potentials. 
The color red, in form of our sensation, seems to not exist neither outside nor 
inside our brain. How do we come to the illusion of the color red? 
In the 
literature this aspect of sensations of perceptions is often denoted 
as qualia \cite{Nagel,Jackson,Block,Dennett}.

In section \ref{bewusstsein} we shall discuss eventually how
we come to the illusion of a ''captain''. It is the same 
question asking for our ''consciousness'', ''self'', or ''I''. Why are we convinced
to have a form of ''I''? And what is the true meaning
of ''consciousness''? Based on the preceding 
discussion about freedom and about decisions and acts we shall arrive at
an interesting model of mind. The mathematician G. W. Leibnitz has discussed
this questions about consciousness already centuries ago \cite{Leibniz}.
In his {\em monade} 17 he is studying the question about 
consciousness comparing our brain with a mill. 
We want to reconsider this remarkable thought experiment 
and shall try to reveal the Nature of consciousness.

It is our aim to develop a consistent model of mind which provides a principal 
understanding of perceptions, actions and eventually consciousness. 
We shall reveal the illusions we have in mind and the main focus will
be to show how and why these illusions arise.
Let us emphasize, that we are looking for a basic model of mind,
which will not consider all the interesting details. For example, when we discuss
our actions, in a general sense also reflexes belong to actions. 
But reflexes occur in a very different manner than conscious actions like
shooting a ball towards a goal. Reflexes follow inevitable on a
certain stimulus; they are ''wired'' firmly  and
they follow therefore a different principle than when we
shoot a ball. But reflexes do not appear to be in contradiction
to our imagination, since we accept them as ''mechanical''.

We will see that our imagination
is very misleading compared to our true nature. 
We shall see, how in our model of mind these illusionary imaginations
arise and shall understand how 
consciousness appears. 
%

\section{Unfree will}
\label{willensunfreiheit}

The central point in the development of a model of mind
is to realize that the freedom of will is an illusion. 
The old quest for freedom is still today 
subject to discussions with very different opinions. 
This quest for freedom was
discussed already by the ancient greeks Democritus \cite{Democritus}
and Aristotele \cite{Aristotle}, newer discussion can be found, for instance,
by 
\cite{Chalmers97,Descartes,Ginet,Locke,Penrose1, Penrose2,Popper,Rosenthal86},
%

In general we are convinced to be masters of our actions and thoughts,
to weight and eventually decide freely. A consequence of this imagination of freedom 
is our understanding of responsibility and guilt, concepts 
which are deeply rooted in our fundamental law;
being guilty of a criminal offense means to commit a violation of criminal law.
Finally, to a more or less large extend, our understanding of good and evil, 
responsibility, guilt and atone are related to religion: 
if we violate God's law we sin and are guilty. 
Despite this big historical heritage with respect to the body-mind problem
we shall argue, that under simple assumptions we never can be free.
First, we have to define what we mean by free.
Let us consider a concrete example:
suppose, a waiter offers two kinds of muffins, 
say a vanilla and a chocolate muffin. 
We choose one of both, say the chocolate muffin. 
If this choice were {\em free} this means that we could,
at the same time, that is, under the exact same conditions,
have made an alternative choice. In this example we mean to 
be free, when we could have chosen the vanilla muffin instead. We preliminary
define freedom as the capability to have made an alternative choice,
under the same conditions. We will see that this definition not quite gives
what we actually expect from freedom; therefore we denote this definition
as preliminary. 
 
Let us note that in praxis it is not possible to restore 
the exact same conditions since this would require to go back in time.
Could we repeat the ''experiment'' of the  
choice of two types of muffins, we would be able to immediately verify
whether we always take the same choice or not. However, we will see that this
is not necessary to understand that we are not free. 

At a first glance, it appears not questionable that we, following
our definition, do have the freedom to choose the alternative vanilla muffin. 
But looking at it again this appears to be impossible, at least 
under certain assumptions:
we suppose that our nervous system follows the same basic principles 
as the rest of Nature does. In detail these principles are electromagnetic and
biochemical interactions in the neurons and its joints, the synapses. 

If we follow {\em natural} processes, then the choice of a muffin is the
result of a cascade of preceding processes. Every process conditions 
the following. When we have made a choice, while we extend our arm
towards the chocolate muffin, then this choice arises from preceding 
(electromagnetic and biochemical) processes. An alternative choice 
means that there were alternative preceding processes present, 
what obviously was not the case. 

This argumentation is the determinism, the principle of
cause and effect. Our choice appears to be determined, 
that is, {\em not} free, because we could not have taken 
another choice under the same conditions or equivalently, at the same 
time. Let us mention that this contradicts our imagination of free will. We 
will come back to this interesting point later.

Firstly, let us consider the assumption that we follow {\em natural} processes. 
Nowadays we undestand the elementary biochemical and electromagnetic processes
in our nervous system. Even that we certainly do not understand our brain 
as a whole, we do understand its basic elementary units, the neurons, together
with its joints, the synapses. Electromagnetic action potentials 
are transmitted through the neurons and in response, neurotransmitters, that is,
signaling molecules, are released and excite the receptors of another neuron. 
Obviously, the basic interactions, that is, its biochemical processes
are principally the same interactions as 
we observe in Nature elsewhere. 

A milestone in this context was the synthetic production of urea. 
Before, the chemistry of life, the organic chemistry, was
strictly separated from the non-organic chemistry of the remaining, {\em dead},
substances. This reflected the idea that chemistry of life is
fundamentally different from that of other substances. With the 
synthetic production of urea in 1828 by W\"ohler \cite{Woehler},
that is, the production of an organic substance from non-organic 
substances, this distinction had to be given up.
The chemists today denote the molecules based on carbon
for historical reasons as {\em organic} chemistry, in contrast
to non-organic chemistry. 
This is certainly a strong evidence supporting our
assumption. 
A further evidence that we follow natural processes 
is that {\em material} influences change our
thoughts and actions. An example of this is the consumption of alcohol. We recognize
the close relation between the material substance alcohol and mind. 
A further example of this close relation
 is given by lesion of certain areals of the brain which result in
mental changes. 
But let us note, that the assumption, that our nervous system
is based exclusively on natural interactions is plausible but difficult to prove. 
We will nevertheless make this assumption in our model of mind and shall see
that we can in deed understand our thoughts and actions in principle. 

Secondly, let us take a closer look at the principle of cause and effect, the
determinism. It reflects our daily experience that every effect has a cause:
if the stone hits the window with sufficient momentum, the glass will break. 
When a  action potential arrives at a synapse, it reveals
a certain amount of neurotransmitters. 
However, we know that this principle of cause and effect has fundamental limitations:
quantum mechanics, discovered by E. Schr\"odinger 
\cite{Schroedinger},
 tells us that under the exact same conditions
we can observe different effects. In Nature there are processes which
occur {\em spontaneously}.
An example of this is the decay of a radioactive element,
which decays spontaneously: if we observe two such radioactive atoms
they in general decay at different times. Before their decay, both
atoms are in every detail identical. 
The reason is not, that we do not know the exact details of
the instable atoms, but merely it is Nature itself following this rule. 
For a sufficient large number of
radioactive atoms we can only give the half-life time, the time 
after which about half of the atoms have decayed. However, for a single atom
we can not know this time of decay, it appears to be undeterminable. 
We can compare this with throwing a coin. When we throw a coin sufficiently
often, we find that about half of them fall on one specific side. 
For a single throw we can not determine the outcome with certainty. 

If we suppose, that our nervous system underlies
the same interactions as everything else we observe in Nature,
then we cannot exclude that there are spontaneous processes. 
With other words, in the cascade of
processes it may happen that there appear processes which are 
not determined. Hence, the processes in the nervous system are natural 
but not necessarily determined!
Following our preliminary definition of freedom, that is, the ability to
make an alternative choice under the same conditions, we appear to be free!

However, we have to realize that this kind of freedom does not satisfy our
idea of freedom. Obviously, considering a machine, employing
a spontaneous mechanism, for instance triggered by radioactive decays
of instable atoms, we would {\em not} call free, even that it 
satisfies our defintion. 
Instead, what we mean by free is to make an alternative choice under the same
conditions but {\em not spontaneously or randomly}. Therefore, let us
 define freedom eventually as the ability to take an alternative choice under the
same conditions but {\em not spontaneously}. 

Accordingly, following this revised definition, our decisions and actions 
are not free, supposed we underly natural processes.

Let us comment on the current discussion about freedom.
The physicist Max Planck
was also engaged in the quest of freedom \cite{Planck1, Planck2}.
He realizes that the spontaneous, random processes 
do not make our actions free, but he tries by a kind of ''inner'' dialog
to declare us free. His argument can be sketched following our example 
of the choice of muffins. 
Suppose the waiter offeres the two types of muffin but we are accompanied 
by a friend. We discuss with our friend the preferences of the two
muffins. Evidently this discussion will influence the process of decision
and this may result eventually in an alternative choice.
Max Planck states that this process of discussion can take place in our
brain in a similar form without our friend present and this makes us free, 
because, following Planck, this may result in an alternative outcome.
But there is a flaw in this argument: of course the friend can affect
our decision and we may even come to an alternative choice.
 However, under the
{\em same} conditions, taking into account our friend, there appears no
alternative process, disregarding for the moment spontaneous processes.
When we think of our friend as a kind of a ''inner'' dialog, 
we get again to an illusory kind of 
''captain''. The ''inner'' dialog in reality is part of the cascade of processes
which never can be free. As we have argued, spontaneously processes do 
not change the argument, because ramdom processes do not count following
our definiton.

In the literature we can find many variants of this argumentation
of Planck; see for instance \cite{Bohm, Penrose1, Penrose2, Hameroff1, Hameroff2, QM}.
Typically there appears a kind of ''captain'' on board of our body in order
to save our freedom. 
The crucial point is to realize the illusion of this kind of ''captain''.

We see, that in order to be free, there appears only the possibility to 
reject the assumption
that mind processes are natural.
We mentioned some evidences which indicate that the interactions in
our nervous system are not anything special, compared to 
 interactions elsewhere,
although it appears to be difficult to prove this.
The argument to look for something beyond our nervous system 
is essentially the believe in a ''soul'' which
is believed to have some kind of existence beyond our body.
Of course, we can not accept this and will instead try
to develop a model of mind under the assumption that mind processes
do not go beyond natural processes elsewhere.

In section \ref{schlusswort} we shall briefly mention
some consequences of the absence of freedom.
In particular it might appear to be unacceptable to be unfree since
this contradicts our imagination. We postpone this
discussion after we discuss in chapter \ref{bewusstsein} how we arrive
at the illusion of consciousness. 

Let us close this section with an illustration by
Carl Ginet \cite{Ginet} who compared our illusion of freedom with a little child
in a ghost train:
the child sits in a small vehicle, equipped with a little unconnected, decorative 
wheel. 
The child is moving the wheel in the illusion to steer
the vehicle which in reality is guided by the rails. 

\section{Thinking and acting}
\label{handeln}

For the moment we want to extend our findings with respect to the lack of freedom
to our thinking. 
So far we have considered the freedom with respect to decisions,
like in the example of the two kinds of muffins. But similar to our 
decisions, thoughts
represent also {\em natural} processes, at least relying on our assumption that 
all processes in our nervous system are of the same nature as processes elsewhere.
Without knowledge about the detailed realization of thoughts in our nervous
system we therefore assume that they represent neuronal processes. 
Strictly speaking it is in this context
irrelevant that thoughts are neuronal processes, it is only
relevant to assume that they are any kind of natural processes.
Then, our argumentation with respect to our decisions 
can directly be applied to our thoughts; thoughts follow from a cascade
of processes, disregarding for the moment spontaneous processes.
Any process of thinking is preceded by a cascade
of other processes, which in turn determine this thinking.

Similar to our definition of freedom of decisions
we mean by free thinking the capability to develop
an alternative thinking under the same conditions, but not in a 
spontaneous way. We see that in analogy to decisions
there is no freedom in our thoughts.

We see how misleading our illusion of a free ''captain'' is. The
freely acting and thinking ''captain'' is to abandon.
Let us emphasize the passivity of the process of thinking: at a closer look
it is not ''us'', who develop this or that thought, but
the thoughts appears in a passive way in our nervous system. 
An active form of thinking, the creation of thoughts, in contrast, 
would correspond to a kind of illusionary ''captain''.
How could a thought arise, if not caused by other processes,
respectively spontaneously?
While I am writing
these lines, it is in fact my nervous system, generating this thoughts -
it is not my autonomous ''I'' in a sense of a ''captain'', who
develops this thoughts. 

In detail, the process of thinking is certainly very complicated, 
for instance, it is affected by experiences and memories. 
These experiences go back probably to our earliest childhood. In addition
we are typically confronted with many perceptions, for instance, a sound which distracts
us. Nevertheless, these details should not obscure the fact 
that thinking is a passive, unfree process. 

We shall look closer at the Nature of thinking in section \ref{bewusstsein},
but we already see, that under the plausible and simple assumption,
that thinking is represented by natural processes, thinking
is as little free as actions and decisions. 

If actions, decisions and thoughts are unfree, the question arises, how 
they are developed instead?
The question is, what instead of a ''captain'' is the principal, unfree
''mechanism''
between the ''input'' given by the monitors, the echo sounder and so forth,
and the output, that is, the steering of the ship. 

Let us fist consider as an example a reflex,
which is in a general sense a kind of action.
For a reflex it is immediate to see the principal unfree ''mechanism'':
when the rubber hammer hits the sensor area at the kneecap then we 
 move our lower leg. 
The evolution has equipped us with reflexes in order to react fast 
and the reflex connects the incoming signals going towards
our nervous system with the outgoing signals, the
motor function given by the muscle contraction.
In the example of a reflex we recognize immediately that the action is not free:
the movement follows inevitably on the stimulus. However, this
action does not contradict our imagination. We accept the reflex as
''mechanical'', in accordance with reality. 

In general, our actions are not reflexes and we can in general adapt our actions
to changing circumstances. 
In case of the choice of the muffin the ''mechanism'' appears to 
be more complicated and this ''mechanism'' is of course not a reflex. 
We weigh the advantages and
disadvantages, have memories, experiences, visual perceptions and many
more aspects which lead to our decision. What is the principle, or 
the ''mechanism'', which has to be passive and cannot be free, in order to
reach the decision?

We propose that our system of desire and pain signals provides the 
fundamental principle. 
The principle, we postulate, is to maximize desire signals and
minimize pain signals. We will see that actions can arise based on this
principle as required in an unfree manner. 
Reflexes are excluded from this principle, as discussed already. 
Considering once again the choice of muffins, the eventual choice 
corresponds to stronger desire signals: our experiences, memories, 
the visual impression and so forth guide our nervous system
to the choice, because it is accompanied by stronger desire signals than
the alternative choice. Maybe, having chosen the chocolate muffin,
we are disappointed, because the taste does not meet our expectations. 
We will memorize this experience and this may lead to an alternative choice in the future. 
Let us note that we are talking about desire signals and pain signals and not about
desire and pain in order to emphasize the ''mechanism''. The sensation
or the experience of desire and pain will be discussed later in section
\ref{wahrnehmung}. To summarize, we postulate that the principal 
''mechanism'' of our nervous system is to reach certain signals 
and to avoid others. 
 
Some remarks are in order: apparently, the principle of maximizing
desire signals and minimizing pain signals is often not immediately obvious. 
However we want to emphasize that it nevertheless may be the basic principle
- excluding reflexes. If we are hungry this is a pain signal which we avoid,
when we eat. We take care of our body, avoid injuries
and other forms of dangers, following this principle.
But if we get up early in the morning and go to work, we can ask,
where we can see this principle?
But of course, we have made the experience that based on this 
habit we keep our job. This is in the long term reflected
by a regular salary and other benefits, which indirectly
correspond to more desire signals. In most cases we do not 
follow this basic ''mechanism'' directly, but by a closer look
we can nevertheless recognize it as a fundamental principle. 
Even when we share our meal with someone, this can be seen as a
gain of desire signals: we have experienced, for instance, that 
it is advantageous for us to share since we expect that if we do so
others will also share with us.

Of course we see the difference between reflexes and other actions 
based on the gain of desire signals 
(together with the avoidance of pain signals). Reflexes are fixed, 
firmly wired, and do not allow to adapt our behavior to changing circumstances. 
We move the lower leg constantly, when the hammer triggers the stimulus. 
But if we have bad experiences with the chocolate muffin, we will probably take
an alternative choice. Learning as a change of behavior due to 
experiences is not possible with reflexes, but certainly 
following the principle of desire and pain signals. Both principles have in common
that we can understand them as unfree ''mechanisms'', as required.  

We can illustrate the principle 
of maximizing desire signals and minimizing pain signals 
with a chess programm:
the chess programm calculates different variants of possible 
moves and values 
the different positions reached in memory. It then chooses the movement
corresponding to the highest value.
This is similar to
the choice of the muffin where we ''value'' both possible moves and choose
the muffin corresponding to the highest ''value'', that is, desire signal. 

Let us in this context consider a rat experiment performed by
James Olds and Peter Milner \cite{Olds}. 
In this experiment an electrode was put into a certain areal of the brain 
of a rat. The rat itself can release an electric signal 
to this electrode by pressing a button.
Before, the rat has been trained to use 
another button
which triggers a mechanism such that feed drops into the box of the rat. 
In the experiment the rat presses the button connected to the electrode
continuously. The rat does not consider the other botton to the point of 
exhaustion. We easily understand this based on our principle of desire signals. 
The electrode hits obviously an area of the nervous system triggering
a strong desire signal.

We may argue that this experiment shows the principle in rats 
and not in humans. But considering persons addicted to drugs,
we recognize parallels. These people typically lose their job, 
neglect social relationships, and often have a tendency to crime -
only in order to gain the desire signal, triggered by drugs.
The brain has found a fatal way to maximize desire signals. 
This may explain, by the way, why it is so difficult
to get people addicted to drugs to give up this destructive way. 

There is no contradiction if we consider someone who hurts himself on purpose. 
If the pain signal is over compensated by a desire signal this can be understood
based on our principle. 
Many actions may appear to not follow this principle
on the first sight, but at a closer
inspection we can recognize its underlying mechanism at work. As has been mentioned,
reflexes are excluded from this principle.

Moreover, we have seen that a ''mechanism'' is required in order 
to explain our actions. Which fundamental principle do we have available
except from our system of desire and pain signals? We can also 
ask what is the meaning of this sophisticated system of pain and desire signals 
other than providing a mechanism of assessment?
Hence, it appears to be exactly the required principle to replace
the inner ''captain''. This principle explains 
our actions and decisions in a consistent and unfree, that is passive manner.
To summarize, we  postulate in our model of mind 
that the principle, maximizing desire signals
and minimizing pain signals is the basic principle of actions and decisions apart
from reflexes.

\section{Perception}
\label{wahrnehmung}

Suppose we watch the sunset with its deeply red sky. 
We know that this perception of a color is another illusion:
before the light hits our retina, it is an electromagnetic
wave in a certain range of wavelength. Of course, nothing 
of the electromagnetic wave is red. In the retina, the incoming
wave excites charges to oscillate in specialized cells.
In turn these cells transform the incoming signal into 
an electric action potential \cite{Pribram}. 
The complete remaining processing proceeds in neurons in terms of 
action potentials which seem to have nothing to do with the sensation
of the color red. 
Neither we can understand the sensation relying on an inner ''captain''
who could get the signals presented on a kind of inner screen. As we have seen
in the discussion in section \ref{willensunfreiheit}, this ''captain'' is an illusion. 

Moreover, the stimulus, after being translated into the ''language''
of the nervous system, that is, being available in form of action potentials,
is already decomposed on its way to the cortex and gets to 
different separated areas. Of course, the signals do not converge anywhere
but are processed further. 

What is then our perception of the color red?
We realize how difficult this is to answer, if
we try to explain the color red to a blind person (someone
who was born blind so that he/she has never experienced this sensation).
It appears to be impossible. 
This problematic can be extended to other sensation in an analogous way, 
and we see that
all our perceptions
appear to be illusions.

The guiding principle to reveal the nature of perceptions is
the finding that this process is required to be passive and can not be free.
Free or active would mean that the perception is ''internally'' represented to 
a kind of ''captain'' on a kind of screen. 
Suppose there would be an inner representation, then, this
representation would have to be watched by some kind of ''inner eye''
and we arrive at 
a senseless loop, also known as infinite regress~\cite{Rosenthal86}. 
Since the perception has
to be a passive process we have to replace the
''captain'' by a passive ''mechanism''.

Of course we know that the meaning of perception is to adapt our behavior to
the surrounding. With our findings in the last section
we know that perceptions serve
to provide informations such that our system of 
gaining desire signals and avoiding pain signals generates
actions and decisions. 
Hence, we have to consistently explain perceptions 
satisfying the following requirements:

\begin{itemize}
\item Perceptions cannot be any form of ''inner'' representation.
\item Perceptions have to be a passive process.
\item Perceptions have to satisfy the functionality to 
get informations, eventually in order to adapt our behavior. 
\end{itemize}

If we ask ourselves what is red, we would describe this perception 
in the following or similar way: the color red is the color of 
an apple, of the sunset, of the ember of fire, our blood and
we think of red when we listen to the sound {\em red} or read the word {\em red}.
Obviously we find that we associate the perception with a bunch of
other sensations. We immediately see that these associations indeed
satisfy all the mentioned requirements. 
Therefore we postulate that these associations {\em are}
the perception which is triggered by the initial stimulus. 
Building associations to the stimulus given by red light, 
occurs without any kind of inner representation,  
is a passive process, and provide us with information about our surrounding.

Let us now consider an auditory perception, when
we for example press the key of a piano keyboard. 
In an analogous way to the visual perception, the sound waves
are nothing but fluctuations of pressure in the air which are
transformed into action potentials in the hearing. 
What is then in a passive form the sensation when we listen
to the sound of a piano?

We associate the auditory signal with a bunch of sensations, for instance
the visual impression of a piano, piano music in our memory, the
visual perception of a concert hall and certainly much more. The whole bunch
of associations, triggered by the initial stimulus is, as we postulate,
the sensation of the piano sound.

A blind person, without memories about visual sensations,
has never experienced the color sensation red and is
therefore not able to build associations. 
Of course, we
see, that the perceptions are individually different and in particular
are influenced by culture. Accordingly,
we expect that the Inuits in Greenland 
have certainly another sensation of the color white than
someone how grew up closer to the equator. 

A little child, told by his parents, ''The apple is red'', 
''This toy is red'' learns the perception, triggered by the 
initial stimulus of light of a certain range of wavelength by
the association to the sound of the spoken word ''red''. The child 
does accordingly {\em not} learn to recognize the color red
on a kind of screen, but learns the perception itself, building
associations.

We should mention that our system of perceptions is remarkable
sophisticated and typically gives very extended associations. But we
are interested in the basic principle here without considering all the details.

Eventually, let us consider the sensation of pain, for instance,
when we accidentally cut our skin with a knife. In an analogous way
the stimulus, triggered by specialized receptor cells 
in the skin, generates action potentials
which are transferred to our nervous system. But there is a principal difference
between these signals referring to pain, and
the visual perception for instance. 
The difference is, that our nervous system, stimulated by the cut, 
tries to avoid this kind of signals. 
As we have seen in the last section, our actions are
driven by the passive principle to avoid pain signals. This distinguishes 
desire and pain sensations from all other perceptions, which are not a part
of our assessment system.

How do we explain the visual perception, when we consider a landscape?
We have discussed already, that the landscape does not appear in form of
an ''inner'' representation. When we watch the landscape, we actually 
recognize different details, a birch there or a cloud above and a lodge over there.
We associate different visual stimulus' entering our eye from different directions
with sounds like ''birch'', or ''lodge''. All together we associate the 
perception maybe with ''valley'', but for this to happen,
different details have to appear from different directions. In fact we are not aware 
of many details, say, a horse, which has been there all the time but 
only through its whinnies got our attention and now compounds to our sensation.
The  ''picture'', that is, the bunch of associations, has changed in this moment, even
that the visual stimulus has not. Our visual system is able to 
distinguish different directions and locations and to recognize patterns. 
We note that the visual system is very advanced. This
is reflected by the fact that the visual system in our cortex occupies
a large part. If we consider the photo of a landscape, this photo 
does not show our ''inner'' representation, but the photo triggers
a sensation which is similar to the landscape itself. 
We therefore associate the landscape with a ''picture'' of it. 

\section{Consciousness}
\label{bewusstsein}

Let us start the discussion of consciousness following
a thought experiment by G. W. Leibniz from his monade 17; see for instance
\cite{Leibniz}: \\

\begin{center}
\begin{em}
\noindent
Besides, it must be admitted that perception, and anything that depends on it,\\
cannot be explained in terms of mechanistic causation -- that is, \\
in terms of shapes and motions. \\
Let us pretend that there was a machine, which was constructed in \\
such a way as to give rise to thinking, sensing, and having perceptions. \\
You could imagine it expanded in size (while retaining the same proportions), \\
so that you could go inside it, like going into a mill. \\
On this assumption, your tour inside it would show you \\
the working parts pushing each other, but never anything which \\
would explain a perception. So perception is to be sought, \\
not in compounds (or machines), but in simple substances. \\
Furthermore, there is nothing to be found in simple substances, \\
apart from perceptions and their changes. \\
Again, all the internal actions of simple \\
substances can consist in nothing other than perceptions and their changes.
\end{em}
\end{center}

We would like to reconsider Leibniz thought experiment,
presented about three centuries ago.
Today we know, that at a {\em tour inside}
the elementary {\em working parts} are
the neurons, which are {\em pushing each other} by means of
electrical and biochemical activity.
Following Leibniz closely it is evident that
we can not find at any special location
anything from 
which we could explain perception, sensing or
thinking. 
This illusionary special location of perception, sensing
or thinking we have denoted as a ''captain'' earlier.
The illusion of a ''captain'' corresponds to our imagination of 
what we would call {\em consciousness}
 or {\em self} or {\em I}.

We have seen, that any special location of
perception, sensing and thinking leads to contradictions:
as Leibniz argues, suppose, we could detect a special location, 
then, in a further expansion
in size we could again go inside and would only find
{\em working parts}, {\em pushing each other}.
Indeed, in terms of neurons, 
we know that the neuronal signals
do not converge anywhere. 
Besides, as we have seen in section \ref{wahrnehmung},
any kind of convergence at some location of
perception would require some new kind of
inner ''eye''.

Leibniz discusses two solutions, to understand consciousness:
firstly, looking for consciousness in the ''simple substance'' itself, that is, 
from a modern point of view, in the neurons itselves.
However, we know that the fundamental function of 
the neurons is to transmit action potentials. This is
consistent with our assumption we have made in section \ref{willensunfreiheit}
that the interactions in the nervous system, based on 
electromagnetic and biochemical processes, are in principle
 not different from Nature elsewhere. 
Hence, we do not agree with the identification of the 
''simple substance'' to be the location of consciousness. 
But let us mention that nevertheless there are attempts,
following Leibniz, to understand consciousness in the
neurons itself; see for instance 
\cite{Penrose1, Penrose2, Hameroff1, Hameroff2, Hameroff3}.

The second possibility, as Leibniz mentions,
is to understand consciousness from the compound.
Contrary to Leibniz we want to follow this way, and try
to undestand consciousness like the phenomena 
of perceptions, sensing, and 
thinking from the
interplay of the neurons. 

Realizing that perception, sensing, and thinking
appear from the cascade of neuronal processes in
an unfree manner we talk about the 
{\em emergence} of these phenomena.
Using the expression {\em emergence} we emphasize 
that perceptions, sensing, and thinkings are 
as required passive processes.

Let us think about this point further.
Imagine, under anesthetic, one neuron after the other would
be replaced by an exact copy. Of course,
in practise this is not possible, but let us consider 
this as a thought experiment.
Since no neuron would be a special location
of perception, sensing or thinking,
we would in no step replace this special location
simply because this location does not exist. 
The essential point is to see that every neuron is
nothing more then a ''mechanical'' device which is
replaced by an equivalent one. 

After recovering from anesthesia {\em we}
would not recognize any change. The neurons
would interact in the same manner as before
and our perception, sensing, and thinking
would appear in the same way.
Here we see clearly the illusion of our 
imagination of {\em we}. Following our misleading
imagination we would expect that at a certain point
our {\em we} would have been removed, that is, the 
illusionary ''captain'' we expect has left the ship. In reality,
there is no ''captain'' who could leave.

Of course, it makes no difference whether
we replace the neurons one by one,
or all at once. Evidently, this means that,
replaced by a copy, {\em we} would develop
perception, sensing, and thinking in the same way!
Hence, suppose that under anesthetic
our body is replaced by a copy, nothing like {\em I} or
{\em consciousness} or {\em self} would be lost. That is, our perception,
sensing, and thinking is not attached
to certain neurons, but appear from
their processes. The emergence of our thinking, sensing
and actions is clearly seen as originating form
the interplay of neurons in this thought experiment.

Let us further imagine that we replace
each neuron in turn by an electronic device, which
replicates exactly the functionality
of the original neuron.
As before we would not remove in any step
a location of perception, sensing or thinking.
In this way {\em we} eventually would be replaced by a machine
under anesthetic and this machine would
develop the same perception, sensing and
thinking and {\em we} could not feel any difference!
In the circuit there would emerge the same processes as before -
supposed the electronic devices 
work like the original neurons.
We recognize that our imagination 
is contrary to the emergence
of perception, sensing and thinking.
We are convinced to have some kind
of {\em I} -- a location where perception,
sensing, and thinking is formed. 
Why are we subject to this illusion?\\
The question is how do we come to this illusion
of consciousness or ''self'' or ''I'' \cite{Maniatis}?
In order to understand this, let us 
see how the
{\em I} appears in our thoughts.
To this end let us consider an example of a perception,
for instance the smell of an apple.
If we communicate to someone this perception,
we say, for instance: ``I smell the scent
of a fresh apple''. We 
use grammatical first-person in order
to communicate our own perception,
distinguishing it from a perception
of someone else. In contrast, with
the communication of ''She smells the scent
of a fresh apple'' we use grammatically third-person
in order to denote the sensation of a third person.

But what happens if we do not communicate
this statement but only realize the smell?
As we have discussed, this thinking must be emergent,
that is, occur in a passive manner. 
We postulate now that 
thinking is nothing but {\em silent}
communication. Thinking then represents
a communication actually directed to another person. When we 
smell the scent of a fresh apple, then
we associate the sensation with the {\em silent} communication
''I smell the scent of a fresh apple''.

First of all we see, that thinking in this form 
occurs in a passive way, as required by our findings.
We further see, that we have to use 
grammatical first-person in the thought. 
We silently communicate our
sensation and not the sensation of someone else. 
The first person ''I'' appears inevitably in our thought. 
This ''I'' or ''self'' is the same as our consciousness.

In this way the illusion of a
location of perception, sensing,
and thinking appears automatically -
a machine would develop
the same illusion of a location
of perception, sensing, and thinking. We understand
now what would happen in the copy of our nervous system
in terms of an equivalent electrical device:
in the copy would
in a passive manner emerge the same illusion of ''I''.
Our copy would be convinced to be conscious,
it would associate the same silent communication using
grammatical first-person -
compare with the ``zombie'' in~\cite{Chalmers97}.
The machine would be equivalent, and just as litte
a ''zombie'' as we - or with equal right we would be
as much a ''zombie'' as the machine. 

Now we can easily understand what it means
to be aware of something: being aware of the scent of
a fresh apple means that there emerge associations
to the silent communication. If we are conscious of a sensation
we communicate it, but not necessarily verbalize this association. 

Let us consider another example, for instance the haptic sensation at 
our soles of our feet. Before we have read these lines, we were probably 
not aware of this sensation. This we can now understand easily. Triggered by the
words written here, in particular ''haptic sensation'' and ''soles'' we associate
the perception with the silent communication of the form: ''I feel 
the ground at the soles of my feet''.
Since it is a silent communication, we use grammatical first-person and
it appears the illusion of an {\em I}. Before we have read these lines, we were 
not aware of this sensation, even that the stimulus was constantly there. 
What was missing  was the association with the silent communication. 

If we think about something, we imagine to develop 
these thinkings. We realize that thinking happens to be 
in reality very different: 
thinking  revealed 
as a passive process emerges in form of silent communication -
if ''I'' cannot concentrate, this means that
the emergent thinking does not follow a certain subject. 

Let us emphasize that 
it appears in principal possible to copy our ''I'' on a machine.
What appears in the machine would not be a copy - it would be ourself.
Our perception, sensing, and thinking would 
appear in the same manner in the copy. As we have argued,
our ''I'' would not be lost in the process of copying.
Of course, nowadays
computers are not sufficiently sophisticated to simulate
the tens of billions of neurons, but there are already
attempts - see for instance \cite{bluebrain}.
Let us summarize what he have found: 
our ''self'' or ''I'' is an illusion in the following sense,
it is not the ''I'' that wakes up in the morning, thinks and
feels, but it is passive communications that inevitable
emerges involving the grammatical first-person.

\section{Conclusions}
\label{schlusswort}

We have seen how misleading our imaginations about our mind are:
under the assumption that the processes in our nervous system are
basically the same as in Nature elsewhere we find that 
our freedom is an illusion. We have seen that we have to understand
perceptions and likewise actions and thoughts as passive processes. 
The imagination of a ''captain'' on board of our body has to be abandoned. 
Our behavior, we have argued, follows the principle of maximizing
desire signals and minimizing pain signals. This ''mechanism'' we
have identified replacing our illusionary ''captain''.
We understand consciousness which
appears in a form of silent communication as
a passive process as required. Eventually,
we arrive at a model of mind which 
at a glance  may appear to reduce us to will-less
''machines''. But what if we are ''machines''?

However, we should realize how powerfull 
these ''machines'' are. These machines have
composed the St Matthew Passion (see for instance the discussion in \cite{Alekos}) and are investigating the Universe. 
Artificial machines are far away from these achievements. 
We are made of a
vast amount of neurons, equipped with dedicated sensors and
very complex motor functions. Robots appear to be ridiculous 
compared to us, even that they already play better chess than every human,
recognize speech and can build associations artificially. 

Let us note that as a consequence of our findings,
concepts like responsibility and guilt 
have no meaning if we are not free. How could we be guilty 
or be responsible for something if we do not have a choice?
We obviously have to think these concepts of guilt and responsibility over. 
Suppose someone steals
a bike, then we actually can not blame the thief, but
still we can blame the action itself.

We have seen that thoughts appear in a passive manner
 and in reality, we are not able to create thoughts
in a free way. This could be misunderstood as a kind of compulsive behavior. 
But compulsive behavior refers to a mental disorder which is 
characterized by repetitive actions or thoughts. This is quite
different from the required passivity of thoughts and actions. 
We are not free in our actions and thinking but this is by no means a
compulsive behavior since this is in general not a repetitive
process.

The concept of creativity, in a inspirational sense of thinking,
has certainly to be given up. 
The process of new insights and ideas is merely a synthesis,
which originates from the vast amount of impressions and memories. 
This is the price we have to pay for giving up freedom. The creator
in us would be nothing but the illusionary ''captain''. 

Eventually, there arises an interesting feature: 
replaced by a machine, {\em we} could be become immortal, supposed 
it is possible to exactly simulate tens of billions of neurons.

\section*{Acknowledgement(s)}
Many thanks go to M. Rezgaoui for very fruitful discussions. 


\end{document}